\documentclass[aps,prb, amsmath,amssymb,floatfix,12pt]{revtex4-1}
\usepackage{tabularx}
\usepackage{bm}
\usepackage{euscript}
\usepackage{graphicx}
\usepackage{color}
\usepackage{amsfonts}
\usepackage{exscale}
\usepackage{amsbsy}
\usepackage{subfigure}
\usepackage{textcomp}
\usepackage{comment}
\usepackage{hyperref}

\numberwithin{equation}{section}

\newcommand{\be}{\begin{equation}}
\newcommand{\ee}{\end{equation}}
\newcommand{\bea}{\begin{eqnarray}}
\newcommand{\eea}{\end{eqnarray}}

\begin{document}

\bigskip
\title{Particle-vortex symmetric liquid}

\bigskip
\author{Michael Mulligan}
\affiliation{Stanford Institute for Theoretical Physics, Stanford University, Stanford, California 94305, USA}
\affiliation{Department of Physics and Astronomy, University of California, Riverside, California 92511, USA}

\bigskip

\date{\today}

\bigskip

\begin{abstract}
We introduce an effective theory with manifest particle-vortex symmetry for disordered thin films undergoing a magnetic field-tuned superconductor-insulator transition.
The theory may enable one to access both the critical properties of the strong-disorder limit, which has recently been confirmed by Breznay et al. [PNAS {\bf 113}, 280 (2016)] to exhibit particle-vortex symmetric electrical response, and the nearby metallic phase discovered earlier by Mason and Kapitulnik [Phys. Rev. Lett. {\bf 82}, 5341 (1999)] in less disordered samples.
Within the effective theory, the Cooper-pair and field-induced vortex degrees of freedom are simultaneously incorporated into an electrically-neutral Dirac fermion minimally coupled to an (emergent) Chern-Simons gauge field.
A derivation of the theory follows upon mapping the superconductor-insulator transition to the integer quantum Hall plateau transition and the subsequent use of Son's particle-hole symmetric composite Fermi liquid.  
Remarkably, particle-vortex symmetric response does not require the introduction of disorder; rather, it results when the Dirac fermions exhibit vanishing Hall effect.
The theory predicts approximately equal (diagonal) thermopower and Nernst signal with a deviation parameterized by the measured electrical Hall response at the symmetric point.
\end{abstract}

\maketitle

\tableofcontents

\section{Introduction}
\subsection{Background}

The magnetic field-tuned superconductor-insulator transition (SIT) in two-dimensional disordered films is a fascinating example of a quantum phase transition [\onlinecite{SondhiGirvinCariniShahar, Goldmanreview2010, SachdevQPT}].
Early ideas [\onlinecite{Fisher1990a},\onlinecite{Fisher1990}] based on scaling near the putative critical point provided a framework for its understanding in terms of ``dirty" (Cooper-pair or field-induced vortex) bosons which undergo a continuous order-disorder transition.
In addition, the possibility that the critical point might exhibit particle-vortex symmetric or ``self-dual"
dc electrical response, 
\begin{align}
\label{resistivitypv}
\rho_{xx}^2 + \rho_{xy}^2 = \Big({h \over 4 e^2} \Big)^2,
\end{align} 
as the temperature $T \rightarrow 0$ was suggested [\onlinecite{Fisher1990a},\onlinecite{Fisher1990}].\footnote{We assume $SO(2)$ rotation symmetry so that the electrical resistivity tensor obeys $\rho_{xx} = \rho_{yy}$ and $\rho_{xy} = - \rho_{yx}$; 
more generally, particle-vortex symmetry requires ${\rm det}(\rho) = (h/4 e^2)^2$.}
If realized, Eq. (\ref{resistivitypv}) is a profound relation between the dissipative response and the measured Hall effect. 

The pioneering experiment [\onlinecite{Hebard1990}] reported a critical (longitudinal) resistance slightly lower than the quantum of Cooper-pair resistance $R_Q = h/4e^2 \simeq 6.45k \Omega/ \square$.
Subsequent measurements [\onlinecite{PaalanenHebardRuel1992,Yazdani1995, Markovic1999, Steiner2008, Ovadia:2013aa, Breznay2016}] have charted a phase diagram in which the most disordered samples (as quantified by the zero field ``normal state" resistance) have a critical resistance equal to $R_Q$ with vanishingly small Hall effect, thereby indicating the possible experimental realization of a self-dual transition, i.e., a disordered critical point at which the Cooper-pair and vortex dynamics are the same.\footnote{More precisely, the universal or infrared dynamics is the same up to a reversal of the background magnetic flux experienced by the Cooper pairs and field-induced vortices; by convention, we take the Cooper pairs to be immersed in positive flux and so duality\cite{fisher1989} dictates that the vortices feel a negative field of magnitude equal to the effective Cooper pair density.}

\subsection{The challenge}

Despite the many successes of prior work [\onlinecite{Fisher1990a, Fisher1990, LutkenRoss1992, Kivelson1992, LutkenRoss1993, DykhneRuzin1994, FradkinKivelson1996, ShimshoniSondhiShahar1997, Pryadko1997, Herbut1998, BurgessDolan2000, BurgessDolan2001}], an explicit theoretical description of a particle-vortex symmetric SIT is lacking.
A natural starting point -- one that we use in Sec. \ref{argumentone} -- is the Landau-Ginzburg theory for Cooper-pairs in which superconductivity is destroyed by the applied magnetic field [\onlinecite{Fisher1990a}].
If the transition is to be self-dual, particle-vortex symmetry requires that the field-induced vortices have the same description.
While a duality transformation [\onlinecite{fisher1989}], in principle, allows us to check whether or not this is the case, it is a challenge to provide an explicit description of the actual critical point.

One reason is that particle-vortex symmetric response Eq. (\ref{resistivitypv}) sets an upper bound on the longitudinal conductivity at the transition: $\sigma_{xx} \leq {4 e^2 \over h}$.
Therefore, a natural description of a self-dual transition would appear to necessitate an exact description of a disordered critical point.

A second difficulty stems from the differing ways in which Cooper pairs and vortices interact. 
Cooper pairs minimally couple to the 3+1-dimensional electromagnetic field, while the vortices instead directly couple to a 2+1-dimensional emergent $U(1)$ gauge field.
How can a self-dual limit arise from theories in which the effective degrees of freedom have such fundamentally different couplings and interactions?

\subsection{The proposal and its application}

The purpose of this paper is to introduce an effective theory -- the {\it particle-vortex symmetric liquid} -- that overcomes these challenges.
Instead of working directly with the Cooper pairs or vortices, we instead make use of an alternative description in terms of electromagnetically neutral ``composite particles" with fermionic statistics.
These composite particles may each be viewed as a bound state of a Cooper pair and field-induced vortex which interact via an emergent Chern-Simons gauge field.
In a sense, particle-vortex symmetry is achieved from the simultaneous incorporation of both degrees of freedom in the effective theory.

Our proposal most directly derives from two recent works [\onlinecite{MulliganRaghuCFsatSIT},\onlinecite{Son2015}].
The experimental observations of self-dual electrical transport [\onlinecite{PaalanenHebardRuel1992,Yazdani1995, Steiner2008, Ovadia:2013aa, Breznay2016}] and simple estimates [\onlinecite{Abrikosovbook}] of the {\it effective} Cooper-pair density in the pertinent materials imply that the Cooper pairs (and field-induced vortices) are at unit filling fraction $\nu = 1$ in the neighborhood of the SIT [\onlinecite{MulliganRaghuCFsatSIT}].
Consequently, a dual description 
in which the critical bosons in non-zero field are traded for ``composite particles" in vanishing effective flux becomes natural and the one that we adopt.
These {\it composite Cooper pairs} or {\it composite vortices} are fermions and enjoy a Fermi liquid-like mean-field description [\onlinecite{PasquierHaldane, Read1998, alicea2005, MulliganRaghuCFsatSIT}].
They are close cousins of the composite fermions [\onlinecite{jain1989, zhang1989, lopezfradkin91, kalmeyer1992, halperin1993}] that have enabled a successful understanding of many aspects of the two-dimensional electron gas (2DEG) in the quantum Hall regime (see Refs. [\onlinecite{jainCF, Fradkinbook, simon1998}] for excellent reviews). 

The precise theoretical incarnation (given in Sec. \ref{proposal}) that the composite bosons/vortices take follows immediately upon combining the recent advance by Son [\onlinecite{Son2015}] (and related works [\onlinecite{WangSenthilfirst2015, maxashvin2015, KMTW2015, WangSenthilsecond2016, Geraedtsetal2015, MurthyShankar2016halfull, mrossaliceamotrunich2015, MulliganRaghuFisher2016, BalramJain16}]) of a composite fermion theory with manifest particle-hole symmetry for the half-filled Landau level and the seminal observation in [\onlinecite{Kivelson1992}] that particle-vortex symmetry in the context of the SIT is mapped to particle-hole symmetry at the integer quantum Hall plateau transition (IQHT).

\begin{figure}[h!]
  \centering
\includegraphics[width=.7\linewidth]{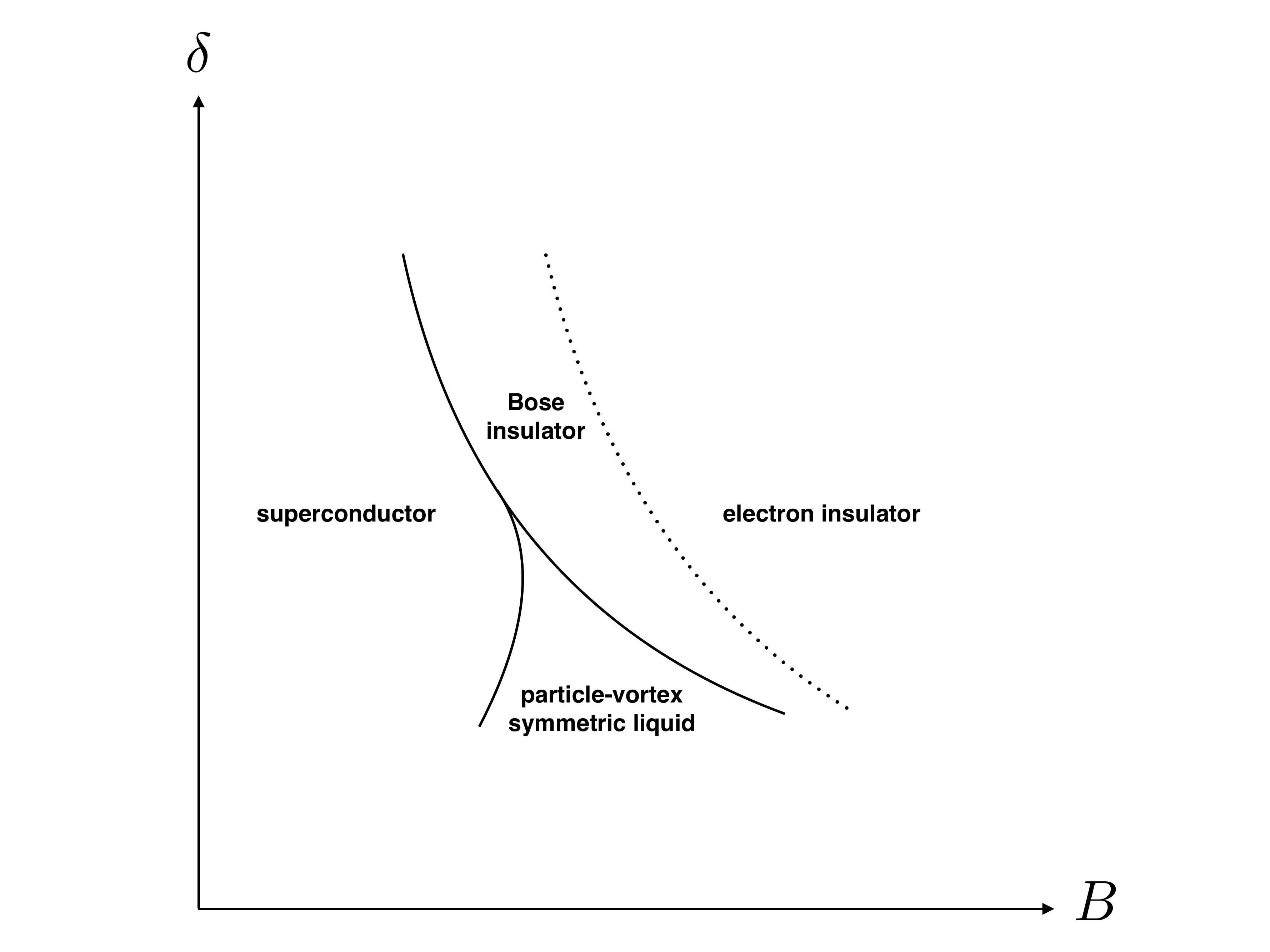}\\
\caption{Schematic $T=0$ phase diagram in the vicinity of the SIT as a function of external magnetic field $B$ and disorder strength $\delta$.  
Solid lines denote phase transitions, while the dashed line signifies the boundary (either transition or crossover) between a Bose insulator and an electron insulator.}
\label{pd}
\end{figure}

In Fig. \ref{pd}, we draw a schematic zero-temperature phase diagram that captures the qualitative, experimetnally-inferred behavior of the disordered films of interest.\footnote{
A similar phase diagram obtains for a 2DEG with the relabeling: superconductor $\leftrightarrow$ integer quantum Hall effect, Bose insulator $\leftrightarrow$ Hall insulator, and particle-vortex symmetric liquid $\leftrightarrow$ particle-hole symmetric liquid.}
We anticipate the regime of validity of the particle-vortex symmetric liquid to lie near the intersection of the superconducting, (Bose) insulating [\onlinecite{Fisher1990a}, \onlinecite{PaalanenHebardRuel1992}, \onlinecite{Crane2007}], and metallic phases and hope it will help illuminate the physics underlying the nearby insulating
phase [\onlinecite{Sacepe2015}, \onlinecite{Ovadia:2015aa}, \onlinecite{Breznay2016}].
The phenomenological utility of our and Son's proposals is to not only provide a novel starting point from which to explain the character of the observed field-tuned SIT and IQHT [\onlinecite{SondhiGirvinCariniShahar}], but also the putative metallic phase that emerges in less disordered samples in the two systems [\onlinecite{Yazdani1995, MasonKapitulnik1999, Kapitulnik2001, Tsen:2016aa, PhillipsBoseMetal2016, Jiang1989transportanomalies, jainCF}].
The observations of metallic phases arising in the vicinity of disordered critical points with an emergent particle-vortex or particle-hole symmetry provide both fascinating and powerful guidance on any putative description of these systems. 

In the limit of strong disorder, we suggest that the particle-vortex symmetric liquid flows to a strong-disorder critical point exhibiting self-dual response.
As the disorder potential is weakened, the Fermi surface of the excitations of the particle-vortex symmetric liquid becomes better defined and the metallic phase emerges. 
Perhaps surprisingly, the particle-vortex symmetric liquid does not require the explicit introduction of a mechanism of composite boson/vortex current relaxation: particle-vortex symmetric electrical response results when the composite bosons/vortices of the theory exhibit vanishing Hall effect.

The remainder of the paper is organized as follows.
In Sec. \ref{proposal}, we introduce the particle-vortex symmetric liquid and summarize a few of its properties, including its expected dc electrical and thermoelectric response.
In Sec. \ref{arguments}, we provide two arguments to derive the effective theory.
We conclude in Sec. \ref{conclusion} and outline possible topics for future study.
Appendix \ref{CVLboundaryderivation} contains the derivation of an equation used in the main text;
Appendix \ref{nonrellimit} sketches the non-relativistic limit of the mass-deformed particle-vortex symmetric liquid; Appendix \ref{WF} discusses the Wiedemann-Franz relation.

\section{Particle-vortex symmetric liquid}
\label{proposal}

We begin with the presentation of the particle-vortex symmetric liquid and then describe a few of its expected properties.
Two arguments that motivate the effective theory are given later in Sec. \ref{arguments}.

\subsection{The effective lagrangian}
\label{pvsymproposal}

The particle-vortex symmetric liquid is described by the lagrangian,
\begin{align}
\label{pvsym}
{\cal L}_{\rm pv} = \bar{\psi} i \gamma^\mu D_\mu \psi + {e_\ast^2 \over 4 \pi} \epsilon^{\mu \nu \rho} \Big( {1 \over 2} \alpha_\mu \partial_\nu \alpha_\rho - 2 A_\mu \partial_\nu \alpha_\rho + A_\mu \partial_\nu A_\rho \Big).
\end{align}
In Eq. (\ref{pvsym}), $\psi$ is an electrically-neutral 2-component Dirac fermion that represents the gapless Cooper-pair boson/vortex excitations in the neighborhood of the SIT; $\alpha_\mu$ with $\mu = t,x,y$ is an emergent gauge field; $A_\mu$ is the external electromagnetic field with background value $\langle \partial_x A_y - \partial_y A_x \rangle \equiv B > 0$ (its third $A_z$ component is ignored in our treatment here).
For convenience, we set $\hbar = 1$, but retain the charge $e_\ast \equiv 2 e$.
The covariant derivative $D_\mu \equiv \partial_\mu - i e_\ast \alpha_\mu$, $\bar{\psi} \equiv \psi^\dagger \gamma^t$, and we take $\epsilon^{txy} = 1$.
The $\gamma$-matrices satisfy the algebra $\{\gamma^\mu, \gamma^\nu \} = 2 \eta^{\mu \nu}$ with $\eta = {\rm diag}(1, -1, -1)$. 
We refer to $\psi$ as the {\it self-dual dyon} (or sometimes dyon for short). 

It is to be understood that dyon self-interactions and couplings to any background potentials may supplement ${\cal L}_{{\rm pv}}$; they remain unspecified in our treatment here.
The former can arise by including the fluctuations of the electromagnetic field where the range of the interaction is dictated by the form of the photon propagator.
Since the electromagnetic field only couples directly to the emergent gauge field, the leading effects on the self-dual dyons come from corrections to the $\alpha_\mu$ propagator and the interaction it mediates.
Quenced chemical potential disorder is incorporated via a background $A_t$ component of the electromagnetic field; it sources vector potential $\alpha_i$ fluctuations of the emergent gauge field.

The electronic charge density and current can be read from Eq. (\ref{pvsym}):
\begin{align}
J^\mu = & {e_\ast^2 \over 2 \pi} \epsilon^{\mu \nu \rho} \partial_\nu (A_\rho - \alpha_\rho).
\end{align}
When the electromagnetic field is taken to be non-dynamical, fluctuations in the electronic density and current are realized as fluctuations of the emergent gauge field, reminiscent of conventional particle-vortex duality.

The $\alpha_t$ equation of motion imposes the constraint:
\begin{align}
e_\ast \psi^\dagger \psi + {e_\ast^2 \over 4 \pi} \epsilon^{tij} \partial_i \alpha_j = {e_\ast^2 \over 2 \pi} \epsilon^{tij} \partial_i A_j.
\end{align}
In the next subsection, we observe that particle-vortex symmetry enforces the solution:
\begin{align}
\langle \psi^\dagger \psi \rangle & = {e_\ast^2 \over 2 \pi} \langle \epsilon^{tij} \partial_i A_j \rangle, \cr
\langle \epsilon^{tij} \partial_i \alpha_j \rangle & = 0.
\end{align}
Thus, unbroken particle-vortex symmetry dictates that the self-dual dyons are placed at a finite density that is fixed by the external magnetic field.
Moving away from unit (bosonic) filling fraction $\nu=1$, the self-dual dyons experience non-zero effective flux as particle-vortex symmetry is explicitly broken. 

In general, the emergent gauge field $\alpha_\mu$ mediates a relevant (in the renormalization group sense) interaction between the self-dual dyons.
While not the focus of this paper, a controlled perturbative study of the clean limit can be set up by introducing a flavor symmetry $\psi \rightarrow \psi_k$ with $k = 1, \ldots, N_f$ [\onlinecite{Polchinski1994}] {\it and} [\onlinecite{SSLee2009OrderofLimits}] either incorporating the effects of a long-ranged Coulomb interaction to soften the $\alpha_\mu$ propagator [\onlinecite{AltshulerIoffeMillis1994, NayakWilczek1994short, NayakWilczeklong1994, Mross2010}] or analytically continuing the theory to $3 - \epsilon$ dimensions [\onlinecite{ChakravartyNortonSyljuasen1995, StanfordGroup2013, TorrobaWang2014}].  
Within a controlled perturbative treatment, the relevant interaction is expected to modify the single-particle self-energy of the self-dual dyons and result in a singular correction to the heat capacity proportional to $T \log(T)$.
These singular corrections supplement the rather dramatic linear in temperature contribution which is expected to result from the existence of the self-dual dyon Fermi sea!

\subsection{Particle-vortex symmetry}
\label{pvsection}

The particle-vortex symmetric liquid is invariant under the combination of the anti-unitary transformation ($i \mapsto - i$)\footnote{The transformation below assumes the $\gamma$-matrix representation $\gamma^t = \sigma^3, \gamma^x = i \sigma^1$, and $\gamma^y = i \sigma^2$ realized in terms of the Pauli-$\sigma$ matrices.},
\begin{align}
\label{pvtransformation}
\psi & \mapsto \gamma^y \psi, \cr
(\alpha_t, \alpha_x, \alpha_y) & \mapsto (\alpha_t, - \alpha_x, - \alpha_y), \cr
(A_t, A_x, A_y) & \mapsto (- A_t, A_x, A_y) + (\alpha_t, - \alpha_x, - \alpha_y), \cr
(t,x,y) & \mapsto (-t, x, y),
\end{align}
and subsequent shift of the lagrangian, 
\begin{align}
\label{shift}
{\cal L}_{\rm pv} \mapsto {\cal L}_{\rm pv} + {e_\ast^2 \over 2 \pi} \epsilon^{\mu \nu \rho} (A_ \mu - \alpha_\mu) \partial_\nu A_\rho.
\end{align}
In terms of the conventionally-defined discrete charge-conjugation ${\cal C}$ and time-reversal ${\cal T}$ symmetries, $\alpha \mapsto {\cal T}(\alpha)$ and $A \mapsto {\cal CT}(A - \alpha)$.
The, perhaps, surprising shifts of the electromagnetic field and lagrangian are explained in Sec. \ref{argumenttwo}.
We identify the combined action in Eqs. (\ref{pvtransformation}) and (\ref{shift}) as the realization of the particle-vortex transformation in the effective theory.

The stability of the particle-vortex symmetric liquid is predicated upon the preservation of the the transformation in Eqs. (\ref{pvtransformation}) and (\ref{shift}).
For instance, a non-zero Dirac mass $\bar{\psi} \psi$ violates the symmetry.
Likewise, unbroken particle-vortex symmetry precludes $\psi$ from realizing a topologically trivial insulator, e.g., via (topologically trivial) Anderson localization, as such a phase is not consistent with the ``parity anomaly" constraint [\onlinecite{NiemiRQ, Redlichparitylong, AlvarezGaumeWitten1984}].
Interestingly, a topologically ordered gapped state analogous to that discovered in Refs. [\onlinecite{BondersonNayakQi, WangPotterSenthil2013, ChenFidkowskiVishwanath2014, MetlitskiKaneFisher2015, MrossEssinAlicea2015}] is allowed by symmetry.

\subsection{Electrical response}
\label{electricalresponse}

One of the most important consequences following from the form of ${\cal L}_{\rm pv}$ comes from the study of its expected dc electrical transport properties by which we can relate electrical response to the self-dual dyon conductivity.
Particle-vortex symmetric electrical response is defined by Eq. (\ref{resistivitypv}).
This condition involves both the dissipative and non-dissipative components of the electrical resistivity tensor.
In this section, we show that Eq. (\ref{resistivitypv}) follows immediately from ${\cal L}_{\rm pv}$ under the natural assumption that the self-dual dyon experiences zero field on average in the neighborhood of the SIT and, therefore, exhibits vanishing Hall effect.
Importantly, no constraint is imposed on the dissipative response of the dyons!

To see this, it is convenient to work in the gauge $\alpha_t = A_t = 0$.
Upon integrating out the self-dual dyons (and terminating the expansion at quadratic order), we obtain
\begin{align}
{\cal L}_{\rm pv} = {i \omega \over 2} {e_\ast^2 \over 2 \pi} \Big( \alpha_j \sigma^\psi_{jk} \alpha_j + \epsilon_{jk} ( {1\over 2} \alpha_j \alpha_k - 2 A_j \alpha_k + A_j A_k)  \Big),
\end{align}
where $\epsilon_{jk} \equiv \epsilon_{tjk}$.
The response of the dyons is captured by the dimensionless conductivity tensor,
\begin{align}
\sigma^\psi_{jk} = \begin{pmatrix}
\sigma_{xx}^\psi & \sigma_{xy}^\psi \cr - \sigma_{xy}^\psi & \sigma^\psi_{xx}
\end{pmatrix}.
\end{align}
The frequency and temperature dependence of the conductivity is left implicit.
Strictly speaking, in order for our intermediate expressions to be well defined, it is necessary to assume an infinitesimal non-zero temperature $T \rightarrow 0$ and a mechanism of self-dual dyon current relaxation so that $\sigma_{xx}^\psi$ is finite\footnote{This latter requirement is readily achieved at finite temperature where an infinite series of irrelevant operators -- at least one of which must surely have a non-zero commutator with the electrical current operator -- are present.}; we shall work under the assumption of finite $\sigma_{xx}^\psi$. 
Integrating out the emergent gauge field $\alpha_\mu$, we find the response lagrangian,
\begin{align}
\label{response}
{\cal L}_{\rm pv} = {i \omega \over 2} {e_\ast^2 \over 2 \pi} A_j \Big(\epsilon_{jk} + (\sigma^\psi + {1 \over 2} \epsilon)^{-1}_{jk}\Big) A_k,
\end{align}
from which we may read off the electrical conductivity.

Using Eq. (\ref{response}), we can calculate the determinant of the electrical resistivity:
\begin{align}
\rho_{xx}^2 + \rho_{xy}^2 = \Big({2 \pi \over e_\ast^2}\Big)^2 {(\sigma_{xx}^\psi)^2 + ({1 \over 2} + \sigma_{xy}^\psi)^2 \over (\sigma_{xx}^\psi)^2 + ({1 \over 2} - \sigma_{xy}^\psi)^2}.
\end{align}
The resistivity has unit determinant (in units of ${2 \pi \over e^2_\ast} \equiv {h \over 4 e^2}$) when the self-dual dyons satisfy the {\it single} constraint, $\sigma_{xy}^\psi = 0$, independent of the value of $\sigma^\psi_{xx}$.
Thus, particle-vortex symmetric response is a consequence of vanishing dyon Hall conductivity in the neighborhood 
where the average magnetic field felt by the dyons is zero.

Although we do not determine the dissipative part $\sigma_{xx}^\psi$ of the self-dual dyon conductivity here,\footnote{This necessitates the introduction of a source of self-dual dyon current relaxation; consequently, we expect a careful treatment of the interplay of disorder and non-Fermi liquid physics to be required.} we observe that a (dimensionless) dyon conductivity $(\sigma_{xx}^\psi, \sigma_{xy}^\psi) = ({1 \over 2}, 0)$ implies the electrical conductivity $(\sigma_{xx}, \sigma_{xy}) = ({e_\ast^2 \over 2 \pi}, 0)$.
The latter is the value of the critical conductivity that has recently been observed at the self-dual SIT [\onlinecite{Steiner2008}, \onlinecite{Breznay2016}].
In general, however, the particle-vortex symmetric liquid allows for a continuous family of self-dual electrical conductivities determined by the value of $\sigma_{xx}^\psi$ (at vanishing $\sigma_{xy}^\psi$).

Within this analysis, the nearby superconducting and insulating phases are easily achieved.
The superconductor occurs when the self-dual (Dirac) dyons exhibit the ``integer" Hall effect with $\sigma_{xy}^\psi = - 1/2$, while the insulator is represented by the Hall effect at $\sigma_{xy}^\psi = + 1/2$ with vanishing longitudinal conductivity $\sigma_{xx}^\psi$ in both phases.

\subsection{Thermoelectric response}
\label{thermoelectric}

We now examine the thermoelectric response of the particle-vortex symmetric liquid following the discussion in [\onlinecite{PotterSerbynVishwanath2015}].
The thermopower and Nernst signal can be extracted from the linear response equation:
\begin{align}
\label{linresponse}
J_i = \sigma_{ij} E^j - \alpha_{ij} \partial^j T.
\end{align}
In Eq. (\ref{linresponse}), $J_i$ is the electrical current, $\sigma_{ij}$ is the electrical conductivity, $E^j$ is the applied electric field, $\alpha_{ij}$ is the thermoelectric coefficient, and $\partial^j T$ denotes a temperature gradient along the $j$-direction.
Using Eq. (\ref{linresponse}), the (diagonal) thermopower $S_{xx}$ and Nernst signal $S_{xy}$:
\begin{align}
\label{thermodef}
S_{xx} & = \sigma^{-1}_{xj} \alpha_{jx}, \cr
S_{xy} & = \sigma^{-1}_{xj} \alpha_{jy},
\end{align}
under the assumption of vanishing electric current (open circuit boundary conditions).

In order to determine $S_{xx}$ and $S_{xy}$, we need to relate $\sigma_{ij}$ and $\alpha_{ij}$ to quantities in the particle-vortex symmetric liquid.
From Sec. \ref{pvsymproposal}, we have the relations,
\begin{align}
\label{dyoncurrentrelations}
j^\psi_i & = {e_\ast^2 \over 4 \pi} \epsilon_{ij} \Big(2 E^j - e^j\Big), \\
\label{electricalcurrentdef}
J_i & = {e_\ast^2 \over 2 \pi} \epsilon_{ij} \Big(E^j - e^j\Big),
\end{align}
obtained from the field equation for the emergent vector potential and the defining relation of the electrical current.
In Eqs. (\ref{dyoncurrentrelations}) and (\ref{electricalcurrentdef}), $j^\psi_i \equiv e_\ast \bar{\psi} \gamma_i \psi$ is the self-dual dyon current and $e^j =  \partial^j \alpha^t - \partial^t \alpha^j$ is the emergent electric field.
Within the particle-vortex symmetric description, a non-zero emergent electric field and temperature gradient result in a linear response relation analogous to Eq. (\ref{linresponse}):
\begin{align}
\label{compositelinresponse}
j^\psi_i = {e_\ast^2 \over 2 \pi} \sigma^\psi_{ij} e^j - \alpha^\psi_{ij} \partial^j T.
\end{align}
(The factor of $e_\ast^2/2\pi$ results from the use of a dimensionless $\sigma_{ij}^\psi$.)
Eq. (\ref{compositelinresponse}) defines the quantities $\sigma_{ij}^\psi$ and $\alpha_{ij}^\psi$.
Equating the expressions for $j^\psi$ in Eqs. (\ref{dyoncurrentrelations}) and (\ref{compositelinresponse}), we  solve for the emergent electric field $e_j$ and substitute into the expression Eq. (\ref{electricalcurrentdef}) for the electrical current to find:
\begin{align}
\label{linresponsedictionary}
J_i = {e_\ast^2 \over 2\pi} \Big(\epsilon_{ij} + (\sigma^\psi + {1 \over 2} \epsilon)^{-1}_{ij}\Big) E^j - {e_\ast^2 \over 2 \pi} \epsilon_{ij} (\sigma^\psi + {1 \over 2} \epsilon)^{-1}_{jk} \alpha^\psi_{kl} \partial^l T.
\end{align}
Comparing Eqs. (\ref{linresponse}) and (\ref{linresponsedictionary}), we can read off $S_{xx}$ and $S_{xy}$ from the definition in Eq. (\ref{thermodef}).

We observed in Sec. \ref{electricalresponse} that particle-vortex symmetry obtains when $\sigma_{xy}^\psi = 0$.
We anticipate that symmetry likewise fixes $\alpha_{ij}^\psi = \alpha^\psi \delta_{ij}$.

It is interesting to express $S_{xx}$ and $S_{xy}$ in terms of the electrical resistivity and $\alpha^\psi$:
\begin{align}
S_{xx} & = \rho_{xx} \alpha^\psi \cr
S_{xy} & = \Big({2 \pi \over e_\ast^2} + \rho_{xy} \Big) \alpha^\psi.
\end{align}
In the vicinity of the experimentally-realized field-tuned SIT  [\onlinecite{Hebard1990, PaalanenHebardRuel1992,Yazdani1995, Steiner2008, Ovadia:2013aa, Breznay2016}], $\rho_{xx} \approx 2\pi/e_\ast^2$ and $\rho_{xy} \approx 0$, and so we expect $S_{xx} \approx S_{xy}$.
This relation can be intuitively understood to reflect the equal contributions from the Cooper-pair and vortex degrees of freedom at a self-dual SIT.

\section{Arguments for the proposal}
\label{arguments}

We now provide two complementary arguments that motivate the particle-vortex symmetric theory.
The first begins within the ordered superconducting phase and uses duality to derive the effective theory.  
The second enlists two different mean-field descriptions of the gapless region near the SIT and uses symmetry to argue for ${\cal L}_{\rm pv}$. 
The agreement between these two approaches gives us confidence in the general proposal.

\subsection{Flux attachment and duality}
\label{argumentone}

The superconducting problem can be described by the effective Landau-Ginzburg lagrangian,
\begin{align}
\label{cooperpairs}
{\cal L}_{\rm SC} = \Phi^\dagger \Big(i \partial_t + e_\ast A_t + {1 \over 2 m_\Phi} (\partial_j - i e_\ast A_j)^2 \Big) \Phi.
\end{align}
$\Phi$ represents the destruction operator of a Cooper pair of effective mass $m_\Phi$ carrying electromagnetic charge $e_\ast$ with respect to the electromagnetism $A_\mu$.
Both here and below, it is understood that additional interactions, consistent with symmetry, are present in the effective lagrangian.
Flux attachment [\onlinecite{zhang1989}, \onlinecite{lopezfradkin91}] posits that the superconducting problem admits a complementary description in terms of a fermion $f$ with lagrangian,
\begin{align}
\label{fluxattachlag}
{\cal L}_f = f^\dagger \Big(i \partial_t + e_\ast (\tilde{a}_t + A_t) + {1 \over 2 m_f} (\partial_j - i e_\ast (\tilde{a}_j + A_j)^2 \Big) f - {e_\ast^2 \over 4 \pi} \epsilon^{\mu \nu \rho} \tilde{a}_\mu \partial_\nu \tilde{a}_\rho.
\end{align}
The emergent gauge field $\tilde{a}_\mu$ statistically transmutes the bosons $\Phi$ into fermions $f$ under the assumption of a classical saddle-point at which the fermions are at unit filling fraction $\langle f^\dagger f \rangle ={e_\ast \over 2 \pi} \langle \partial_x \tilde{a}_y - \partial_y \tilde{a}_x \rangle > 0$ with respect to $\tilde{a}_\mu$ and exhibit the integer quantum Hall effect.

A sufficiently strong external magnetic field $B = \langle \partial_x A_y - \partial_y A_x \rangle > 0$ eventually destroys superconductivity.
Within the fermionic description, the external magnetic field lowers the effective filling fraction and leads to the destruction of the integer quantum Hall effect.
At the point where the Landau level of the fermions is half full, an additional duality transformation enables a description in terms of the particle-hole symmetric composite fermion liquid of Son [\onlinecite{Son2015}]:
\begin{align}
{\cal L}_{\rm pv} = \bar{\psi} i \gamma^\mu D_\mu \psi - {e_\ast^2 \over 4 \pi} \epsilon^{\mu \nu \rho} \Big((\tilde{a}_\mu + A_\mu) \partial_\nu \alpha_\rho - {1 \over 2} (\tilde{a}_\mu + A_\mu) \partial_\nu (\tilde{a}_\rho + A_\rho) + \tilde{a}_\mu \partial_\nu \tilde{a}_\rho \Big),
\end{align}
where the 2-component Dirac fermion $\psi$ is minimally coupled to the emergent gauge field $\alpha_\mu$ through the covariant derivative $D_\mu \equiv \partial_\mu - i e_\ast \alpha_\mu$.
We have implemented the particle-hole symmetric formulation of the composite Fermi liquid in the fermion sector and assumed the statistically-transmuting gauge field $\tilde{a}_\mu$ to remain unaffected.
To simplify the above lagrangian, we integrate out $\tilde{a}_\mu$ and obtain the particle-vortex symmetric liquid in Eq. (\ref{pvsym}).\footnote{
Similar logic applied to the particle-vortex dual of Eq. (\ref{cooperpairs}) yields an effective lagrangian in which $e_\ast \rightarrow - e_\ast$ and the sign of the Chern-Simons term involving the $\alpha_\mu$ field only is reversed.} 

It is interesting to note that the $f$ fermions exhibit the dimensionless conductivity $(\sigma_{xx}^f, \sigma_{xy}^f) = ({1 \over 2}, {1 \over 2})$ (the universal value found at IQHTs [\onlinecite{Shahar1995}, \onlinecite{SondhiGirvinCariniShahar}]) when the self-dual dyon conductivity studied in Sec. \ref{electricalresponse} takes the value $(\sigma_{xx}^\psi, \sigma_{xy}^\psi) = ({1 \over 2}, 0)$.
This is consistent with the observation in [\onlinecite{Kivelson1992}, \onlinecite{ShimshoniSondhiShahar1997}] that particle-vortex symmetric response at a SIT is mapped to particle-hole symmetric response at the IQHT.
The above derivation and the lagrangian in Eq. (\ref{pvsym}) give an explicit realization of this relation.

\subsection{Emergent symmetry restoration}
\label{argumenttwo}

Under the assumption that the effective filling fraction of the Cooper pairs (or vortices) $\nu \sim 1$ in the neighborhood of the field-tuned SIT, it is natural to enlist a mean-field description in terms of ``composite particles" that experience zero flux on average [\onlinecite{MulliganRaghuCFsatSIT}].
However, within perturbation theory, there appears to be two distinct choices: a Fermi liquid-like state of composite Cooper pairs or one of composite vortices.
We will first describe these two effective theories in some detail in order to define a map -- the particle-vortex transformation -- that exchanges them.
We will then argue that these two choices motivate the particle-vortex symmetric liquid as the effective description that obtains in the limit when particle-vortex symmetry is restored.  

\subsubsection{Composite Cooper-pair and composite vortex bulk lagrangians}
\label{compositebulk}

Flux attachment says that Cooper-pair bosons at unit filling fraction can equivalently be described by the lagrangian: 
\begin{align}
\label{CBLbulk}
{\cal L}_{{\rm CBL}} = \psi^\dagger_\Phi \Big(i \partial_t + e_\ast a_t + {1 \over 2 m_{\rm \Phi}} (\partial_j - i e_\ast a_j)^2  \Big) \psi_\Phi + {e_\ast^2 \over 4 \pi} \epsilon^{\mu \nu \rho} (a_\mu - A_\mu) \partial_\nu (a_\rho - A_\mu).
\end{align}
In this {\it composite (Cooper-pair) boson liquid} (CBL), $\psi_\Phi$ is the destruction operator of a composite Cooper pair of effective mass $m_{\rm \Phi}$, $a_\mu$ is an emergent gauge field, and $A_\mu$ again represents electromagnetism with non-zero average magnetic field $B > 0$.
Although closely related, the lagrangians in Eqs. (\ref{CBLbulk}) and (\ref{fluxattachlag}) contain Chern-Simons terms for the emergent gauge fields of opposite level.\footnote{We have presented the CBL lagrangian in a form where the composite Cooper pairs do not minimally couple to the electromagnetic field; a transfer of charge can be accomplished by a field redefinition.
We apply this convention to the composite vortices in the composite vortex lagrangian introduced below.}

Particle-vortex duality [\onlinecite{fisher1989}] allows the field-tuned SIT to alternatively be studied using the induced vortex degrees of freedom.
To use this duality, we implicitly assume that the proximate insulator is a Bose insulator [\onlinecite{Fisher1990a}], i.e., an insulator of localized Cooper pairs -- a possibility that appears to be realized in a variety of materials [\onlinecite{PaalanenHebardRuel1992}, \onlinecite{Crane2007}].
Vortices at unit filling fraction motivate a description in terms of the lagrangian:
\begin{align}
\label{CVLbulk}
{\cal L}_{\rm CVL} = \psi^\dagger_{\rm v} \Big(i \partial_t + e_\ast b_t + {1 \over 2 m_{\rm v}} (\partial_j - i e_\ast b_j )^2  \Big) \psi_{\rm v} - {e_\ast^2 \over 4 \pi} \epsilon^{\mu \nu \rho} \Big(b_\mu \partial_\nu b_\rho - 2 \tilde{A}_\mu \partial_\nu (A_\rho - b_\rho) + \tilde{A}_\mu \partial_\nu \tilde{A}_\rho \Big).
\end{align}
In the {\it composite vortex liquid} (CVL), $\psi_{\rm v}$ is the destruction operator of a composite vortex of effective mass $m_{\rm v}$, $b_\mu$ is responsible for the statistical transmutation of the bosonic vortices into fermionic composite vortices, while $\tilde{A}_\mu$ represents the fluctuations of the Cooper pairs of average effective density $n_s \equiv \langle \partial_x \tilde{A}_y - \partial_y \tilde{A}_x \rangle/\Phi_0$, where $\Phi_0 \equiv hc/e_\ast = 2\pi/e_\ast$ is the magnetic flux quantum.
The average density of composite vortices is fixed by the external field, $n_v \equiv \langle \psi^\dagger_{\rm v} \psi_{\rm v} \rangle = B/\Phi_0$. 
It is convenient to simplify ${\cal L}_{\rm CVL}$ by integrating out $\tilde{A}_\mu$ to find
\begin{align}
\label{CVLbulkreduced}
{\cal L}_{\rm CVL} = \psi^\dagger_{\rm v} \Big(i \partial_t + e_\ast b_t + {1 \over 2 m_{\rm v}} (\partial_j - i e_\ast b_j )^2  \Big) \psi_{\rm v} - {e_\ast^2 \over 4 \pi} \epsilon^{\mu \nu \rho} \Big(b_\mu \partial_\nu b_\rho - (b_\mu - A_\mu) \partial_\nu (b_\rho - A_\rho)\Big).
\end{align}

\subsubsection{Composite Cooper-pair and composite vortex boundary lagrangians}

If the CBL is placed in the lower half-plane ($y < 0$) with topologically trivial vacuum in the upper half-plane ($y > 0$), gauge invariance requires the presence of the boundary degree of freedom $\phi_\Phi$ living at $y = 0$ with lagrangian:
\begin{align}
\label{CBLboundary}
{\cal L}_{\partial {\rm CBL}} = {e_\ast^2 \over 4 \pi} \Big[(\partial_t \phi_\Phi + a_t - A_t) (\partial_x \phi_\Phi + a_x - A_x) - v_{\Phi}(\partial_x \phi_\Phi + a_x - A_x)^2 + \epsilon_{\mu \nu y} (a_\mu - A_\mu) \partial_\nu \phi_\Phi \Big].
\end{align}
Together, the composite boson bulk and boundary theories are invariant under the gauge transformations:
\begin{align}
a_\mu & \mapsto a_\mu + \partial_\mu \Lambda_a, \cr
A_\mu & \mapsto A_\mu + \partial_\mu \Lambda_A, \cr
\psi_\Phi & \mapsto e^{i e_\ast \Lambda_a} \psi_\Phi, \cr
\phi_\Phi & \mapsto \phi_\Phi - (\Lambda_a - \Lambda_A).
\end{align}
The last term in Eq. (\ref{CBLboundary}) cancels the (anomalous) gauge variation of the bulk Chern-Simons term in Eq. (\ref{CBLbulk}).

The operator $\Psi_{\rm B} \equiv e^{i e_\ast \phi_\Phi} \psi_\Phi$ is neutral with respect to the emergent gauge symmetry and carries electromagnetic charge $e_\ast$ (it destroys a left-moving $\phi_\Phi$ mode and composite Cooper pair).
We therefore identify $\Psi_{\rm B}$ with the Cooper-pair boson destruction operator along the boundary.

Analogous considerations (see Appendix \ref{CVLboundaryderivation} for a derivation) result in the boundary lagrangian,
\begin{align}
\label{CVLboundary}
{\cal L}_{\partial {\rm CVL}} & = {e_\ast^2 \over 4 \pi} \Big[(\partial_t \phi_1 + A_t - b_t) (\partial_x \phi_1 + A_x - b_x) - v_{1}(\partial_x \phi_1 + A_x - b_x)^2 + e_\ast \epsilon_{\mu \nu y} (A_\mu - b_\mu) \partial_\nu \phi_1 \cr
& - (\partial_t \phi_2 - b_t) (\partial_x \phi_2 - b_x) - v_{2}(\partial_x \phi_2 - b_x)^2 + \epsilon_{\mu \nu y} b_\mu \partial_\nu \phi_2\Big]\delta(y=0),
\end{align}
when the CVL is placed in the lower half-plane.
The boundary degrees of freedom ensure invariance under the gauge transformations,
\begin{align}
b_\mu & \mapsto b_\mu + \partial_\mu \Lambda_b, \cr
A_\mu & \mapsto A_\mu + \partial_\mu \Lambda_A, \cr
\psi_{\rm v} & \mapsto e^{i e_\ast \Lambda_b} \psi_{\rm v}, \cr
\phi_1 & \mapsto \phi_1 - (\Lambda_A - \Lambda_b), \cr
\phi_2 & \mapsto \phi_2 + \Lambda_b.
\end{align}

Evidently, there are two independent local boundary operators that are neutral with respect to the $b_\mu$ gauge symmetry: $\Psi_1 \equiv e^{- i e_\ast \phi_1} \psi_{\rm v}$ carries electromagnetic charge $e_\ast$ and can be identified with a Cooper-pair boson destruction operator (it creates a left-moving $\phi_1$ mode and destroys a composite vortex); $\Psi_2 \equiv e^{- i e_\ast \phi_2} \psi_{\rm v}$ carries no electromagnetic charge (it destroys a right-moving $\phi_2$ mode and a composite vortex).
In particular, $\phi_2$ and $\psi_{\rm v}$ do not minimally couple to electromagnetism.
$\Psi_2$ can be identified with a vortex destruction operator along the boundary.
The appearance of the boundary field $\phi_1$ is similar to the filled Landau level boundary mode in the composite hole liquid introduced in [\onlinecite{BMF2015}].
The third operator $\Psi_3 \equiv e^{i e_\ast (\phi_2 - \phi_1)}$ is proportional to $\Psi_2^\dagger \Psi_1$ since the composite vortex density is fixed by the external magnetic field.

\subsubsection{Conjugate perturbative descriptions}

The CBL and CVL theories were found by applying the flux attachment procedure to particle-vortex dual descriptions of the SIT.
We now define a mapping under which the CBL and CVL lagrangians
are conjugate.
If we act twice with the transformation, we recover the original lagrangian; this is equivalent to taking the duality transformation interchanging Cooper-pair bosons and vortices to act as an element in $PSL(2,\mathbb{Z})$.\footnote{The modular group appears to be realized differently on non-relativistic and relativistic systems\cite{WittenSL2Z}.}

At the level of the CBL and CVL bulk lagrangians in Eqs. (\ref{CBLbulk}) and (\ref{CVLbulk}), we implement the transformation by combining the anti-unitary mapping ($i \mapsto - i$),
\begin{align}
\label{transformationappendix}
(a_t, a_x, a_y) & \mapsto (a_t, - a_x, - a_y), \cr
(A_t, A_x, A_y) & \mapsto (- \tilde{A}_t, \tilde{A}_x, \tilde{A}_y), \cr
(t, x, y) & \mapsto (- t, x, y),
\end{align}
with the shift of the lagrangian by ${1 \over 2 \pi} \epsilon^{\mu \nu \rho} \tilde{A}_\mu \partial_\nu A_\rho$.
The relabeling $a_\mu \leftrightarrow b_\mu$ and $\psi_\Phi \leftrightarrow \psi_{\rm v}$ completes the transformation under the assumption $m_\Phi = m_{\rm v}$
For ease of reference we refer to this combined transformation by Eq. (\ref{transformationappendix}), however, we emphasize that the full transformation includes both the (local) mapping of fields and lagrangian shift.

It is more convenient to study the CVL lagrangian in Eq. (\ref{CVLbulkreduced}) which was found by choosing the gauge $\tilde{A}_t = 0$ and subsequently integrating out the spatial components $\tilde{A}_i$.
Since $A_\mu$ and $\tilde{A}_\mu$ transform into one another in Eq. (\ref{transformationappendix}), self-consistency requires that we also take $A_t = 0$.
As noted in Appendix \ref{CVLboundaryderivation}, the $\tilde{A}_t = 0$ gauge fixes $\tilde{A}_i = A_i - b_i + \partial_i \phi_1$, which when substituted into Eq. (\ref{CVLbulk}) gives the simplified bulk lagrangian in Eq. (\ref{CVLbulkreduced}) and the contribution to the boundary lagrangian in Eq. (\ref{boundarycontribution}).
No such equation results from fixing $A_t = 0$ since it is taken to be a non-dynamical field.

In $A_t = \tilde{A}_t = 0$ gauge, the transformation in Eq. (\ref{transformationappendix}) becomes
\begin{align}
\label{transformationappendixtwo}
(a_t, a_x, a_y) & \mapsto (a_t, - a_x, - a_y), \cr
(A_x, A_y) & \mapsto (A_x, A_y) - (a_x, a_y) + \partial_i \phi_1, \cr
(t, x, y) & \mapsto (- t, x, y).
\end{align}
Combined with the shift of the lagrangian by ${1 \over 2 \pi} \epsilon^{\mu \nu \rho} \tilde{A}_\mu \partial_\nu A_\rho$ with $(\tilde{A}_t, \tilde{A}_i) = (0, A_i - a_i + \partial_i \phi_1)$ and $A_t = 0$, Eq. (\ref{transformationappendixtwo}) allows us to infer the transformation of the boundary fields,
\begin{align}
\label{edgetransformation}
\phi_\Phi & \mapsto \phi_2 + c_\Phi, \cr
\phi_1 & \mapsto - \phi_1 + c_1, \cr
\phi_2 & \mapsto \phi_\Phi + c_2,
\end{align}
up to a shift by undetermined constants.
We have set the velocities of the edge fields to zero in verifying the conjugacy of the boundary lagrangians under Eq. (\ref{edgetransformation}).
Note that the bulk contributes ${e_\ast^2 \over 4 \pi} (\partial_t \phi_1 - c_t) \partial_x \phi_1$ with $c_t = b_t$ or  $a_t$ to the transformation of the boundary lagrangian. 
We see that Eq. (\ref{transformationappendixtwo}) and subsequent shift of the lagrangian coincide with the action of the particle-vortex transformation on the gauge fields defined in Sec. \ref{pvsection}.


\subsubsection{The argument for restoration}

It is expected that the description of the physics away from $\nu = 1$ in terms of either composite Cooper pairs or vortices is different.
But why do the descriptions in Eqs. (\ref{CBLbulk}) and (\ref{CVLbulkreduced}) and their respective boundary completions appear to differ physically at the putative self-dual point where duality predicts identical physics?
We believe the difficulty lies in perturbation theory about the mean-field saddle-points for the CBL and CVL theories.

To highlight the (perturbative) inadequacy, it is useful to calculate the determinant of the electrical conductivity tensor produced by the mean-field saddles of the CBL and CVL theories in order to test how the self-duality condition in Eq. (\ref{resistivitypv}) might be satisfied.
Assuming finite longitudinal composite boson/vortex conductivity, a mean-field treatment of the two theories results in self-dual electrical response at $\nu=1$ only if the CBL Hall conductivity,
\begin{align}
\label{CBLhall}
\sigma_{xy}^{\rm CBL} = {1 \over 2} {e_\ast^2 \over 2 \pi},
\end{align}
or the CVL Hall conductivity
\begin{align}
\label{CVLhall}
\sigma_{xy}^{\rm CVL} = - {1 \over 2} {e_\ast^2 \over 2 \pi}.
\end{align}
Such a large Hall effect would not be expected at $\nu=1$ where the composite bosons/vortices experience vanishing effective magnetic flux on average.
(Recall that the particle-vortex symmetric theory requires the self-dual dyons to exhibit vanishing Hall effect -- in a sense the ``average" of the above two values.\footnote{One might ask: How trustworthy are these mean-field transport computations? The key difference between the CBL/CVL and particle-vortex symmetric theories is that unbroken particle-vortex symmetry forces $\sigma_{xy}^\psi = 0$, while there is no such constraint on $\sigma_{xy}^{\rm CBL/CVL}$.})
We interpret this ``inconsistency" as a reflection of the inadequacy of perturbation theory about the mean-field saddle-points.
An identical issue arises in the context of the half-filled Landau level [\onlinecite{kivelson1997}, \onlinecite{BMF2015}].

A hint at a possible resolution comes from the mapping defined in Eq. (\ref{transformationappendixtwo}) that transforms the CBL and CVL lagrangians into one another.
The fact that the CBL and CVL theories are not invariant under this mapping (at least within perturbation theory) helps to explain the challenging requirement, highlighted by Eqs. (\ref{CBLhall}) and (\ref{CVLhall}), that self-duality imposes on the CBL and CVL theories.

\begin{figure}[h!]
  \centering
\includegraphics[width=.6\linewidth]{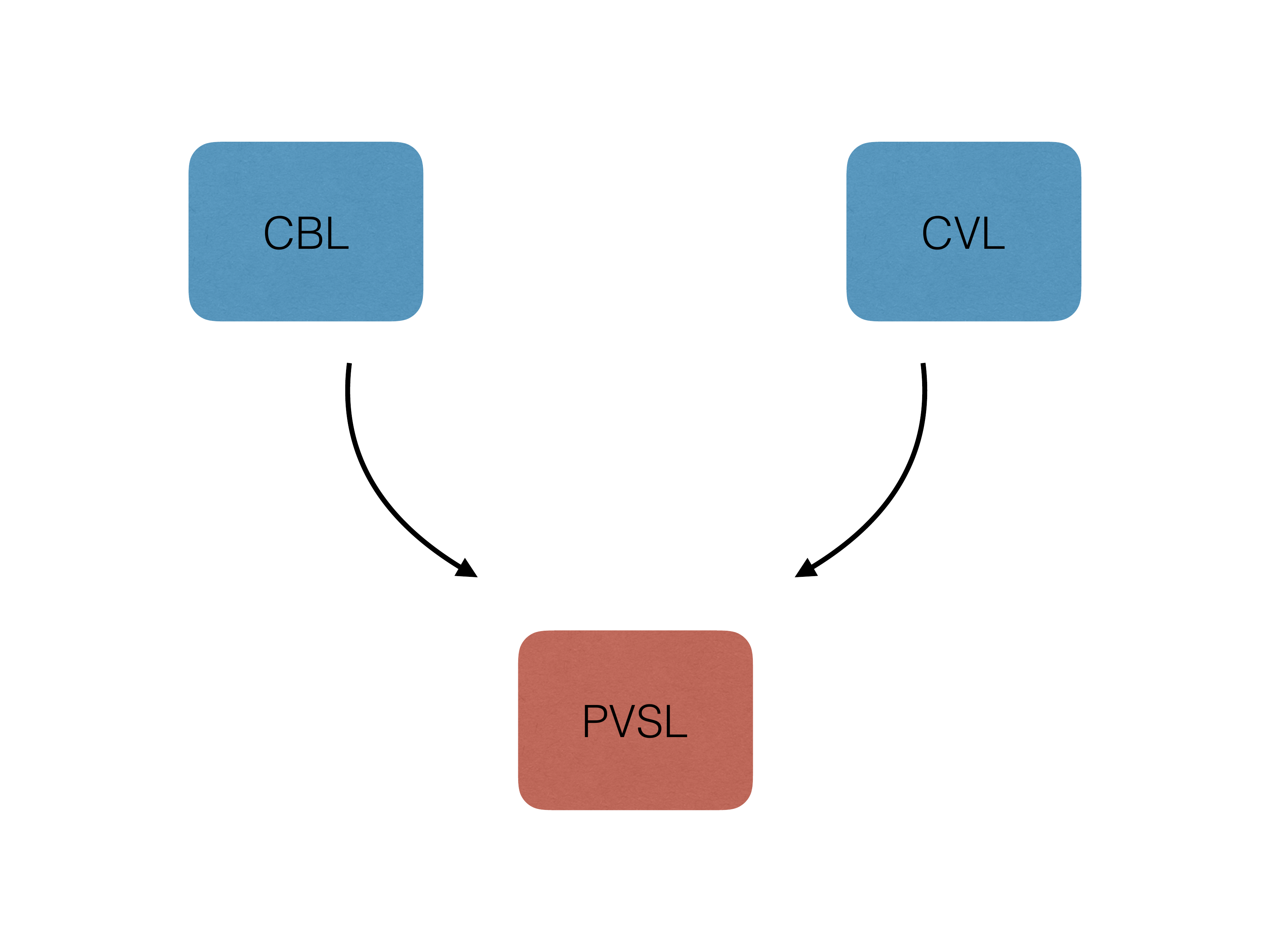}\\
\caption{Particle-vortex symmetry restoration: the composite boson liquid (CBL) and composite vortex liquid (CVL) define distinct perturbative theories; particle-vortex symmetry can be restored by appropriate tuning or possibly renormalization group flow (indicated by the arrows) to the particle-vortex symmetric liquid (PVSL) at long wavelengths.}
\label{rgpicture}
\end{figure}

If the CBL and CVL theories coincide non-perturbatively (or possess a self-dual limit upon variation of appropriate parameters), then we expect the resulting description to be symmetric under the transformation in Eq. (\ref{transformationappendixtwo}) and realized by the particle-vortex symmetric liquid.
The particle-vortex transformation described in Sec. \ref{pvsection} -- under which the particle-vortex symmetric liquid is invariant -- coincides with the transformation in Eq. (\ref{transformationappendixtwo}) up to its action on the matter fields.
The composite bosons and composite vortices are interchanged by Eq. (\ref{transformationappendixtwo}); this may be identified with the exchange of the upper and lower components of the self-dual dyon Dirac spinor under the particle-vortex transformation in Eq. (\ref{pvtransformation}).
In this way, the self-dual dyon incorporates the composite Cooper-pair boson and composite vortex degrees of freedom.
We thus come to the picture of particle-vortex symmetry restoration in Fig. \ref{rgpicture}.

At weak gauge coupling and for non-zero symmetry-breaking mass $m \bar{\psi} \psi$, the particle-vortex symmetric liquid flows to either the CBL or CVL upon taking the non-relativistic limit (see Appendix \ref{nonrellimit}), depending upon the sign of the mass $m$.
It is possible that a large anomalous dimension could alter this weak coupling intuition, driving the Dirac mass irrelevant (in the renormalization group sense) about the particle-vortex symmetric liquid fixed point, and thereby provide additional support for the picture in Fig. \ref{rgpicture}.

\section{Discussion}
\label{conclusion}

In this paper, we introduced the particle-vortex symmetric liquid in order to describe a field-tuned superconductor-insulator transition with approximate particle-vortex symmetry.
In this theory, Cooper-pair bosons or field-induced vortices undergoing an order-disorder transition are traded for an electrically-neutral (composite) Dirac fermion -- that we refer to as a self-dual dyon -- which experiences zero ``magnetic" flux on average and interact via an emergent Chern-Simons gauge field.
In contrast to the composite Fermi liquid treatment of bosons at $\nu=1$, within a mean-field treatment, this theory exhibits particle-vortex symmetric electrical transport, i.e., Eq. (\ref{resistivitypv}), when the self-dual dyons have vanishing Hall effect. 
Furthermore, no constraint need be imposed on the current relaxation mechanism of the dyons; self-dual transport can be satisfied via a single condition!
In addition, we examined the expected thermoelectric response and found $S_{xx} \approx S_{xy}$ in the vicinity of a self-dual superconductor-insulator transition with a deviation parameterized by the (electrical) Hall response.

There are a variety of directions for future exploration.

Perhaps, the most pressing is a careful treatment of the effects of disorder on the particle-vortex symmetric liquid. 
The working hypothesis of this paper is that at strong disorder, the theory flows to a strong-disorder critical point.
A demonstration of the flow and characterization of the disordered critical point are crucial to the picture presented here. 
Can the particle-vortex symmetric liquid provide a correct demonstration of the measured critical exponents?
If a disordered critical point is achieved, what is the value of the critical longitudinal conductivity?
We believe the particle-vortex symmetric liquid may enable a more accessible attack on these problems.

The metallic region (see Fig. \ref{pd}) found to obtain in cleaner samples does not appear to exhibit self-dual response at any value of the applied magnetic field [\onlinecite{Yazdani1995, MasonKapitulnik1999, Kapitulnik2001, Tsen:2016aa}].
Within the context of the particle-vortex symmetric liquid, this violation of self-duality indicates the existence of a non-zero symmetry-breaking mass and a possible effective description of the metallic phase advocated in [\onlinecite{MulliganRaghuCFsatSIT}].
(An alternative proposal of the clean limit might instead utilize the ideas presented in [\onlinecite{Galitski2005}].)

In Sec. \ref{electricalresponse}, we showed how the particle-vortex symmetric liquid achieves self-dual electrical response and described the expected thermoelectric response in Sec. \ref{thermoelectric}.
Is there a thermal transport analog of Eq. (\ref{resistivitypv})?
Furthermore, it is important to examine thermal transport more generally and any signatures of unbroken particle-vortex symmetry.
See Appendix \ref{WF} for a discussion of how the particle-vortex symmetric liquid obeys the Wiedemann-Franz ``law" following the logic in [\onlinecite{WangSenthilsecond2016}].

It would be interesting to better understand whether the particle-vortex symmetric liquid motivates a modification to the wave function advocated to describe bosons at $\nu=1$ [\onlinecite{PasquierHaldane}, \onlinecite{Read1998}]. 
Such wave functions, in the context of the half-filled Landau level, have provided crucial insights to the physics [\onlinecite{jainCF}] and so it is worthwhile to better understand the role that unbroken particle-vortex symmetry (or particle-hole in the half-filled Landau level) may play.

Recent commensurability experiments [\onlinecite{Kamburov2014}, \onlinecite{LiuDengWaltz}] have tested the validity of the composite fermion picture and partially motivated a reconsideration of the conventional theoretical description [\onlinecite{halperin1993}] of the half-filled Landau level [\onlinecite{Son2015}, \onlinecite{BMF2015}].
We believe that similar experiments performed on the thin films could provide valuable information regarding the scenario proposed in Ref. [\onlinecite{MulliganRaghuCFsatSIT}] and extended here in which a Fermi liquid-like state arises from a collection of interacting bosons.

As we have emphasized, our work closely parallels recent studies of the role of emergent particle-hole symmetry in the composite fermion treatment of the half-filled Landau level.
At electronic filling fraction $\nu_e=1/4$, there is no such electronic particle-hole symmetry, however, quarter electronic filling fraction translates to half-filling of the composite fermions.
Might there be an emergent composite fermion particle-hole symmetry pertinent in the vicinity of the fractional quantum Hall transition $\nu_e = 1/3 \rightarrow 0$\footnote{We thank C. Kane for a comment on this topic.}?
Indeed, there is electrical transport evidence that such a symmetry is realized [\onlinecite{Shahar1996}]; commensurability experiments may provide additional insight.

During the completion of this work, there appeared two papers studying the implementation of particle-hole, rather than particle-vortex, symmetry in fermionic and bosonic systems at various filling fractions [\onlinecite{WangSenthil2016},\onlinecite{MrossAliceaMotrunichbosonicph2016}].






\acknowledgments

We thank Matthew Fisher, Steve Kivelson, Chetan Nayak, Sri Raghu, and David Tong for helpful conversations.
We also appreciate the thoughtful comments by Shamit Kachru, Amartya Mitra, and Sri Raghu on an early draft of this paper.
We thank David Tong for sharing with us an early preprint of his paper with Andreas Karch [\onlinecite{Karch:2016sxi}], see also [\onlinecite{Seiberg:2016gmd}], where the zero-field limit of the particle-vortex symmetric lagrangian is shown to be dual to the theory of a Wilson-Fisher boson.
We are grateful for the generous hospitality of the Kavli Institute for Theoretical Physics in Santa Barbara where this work was completed.
This research was supported in part by the John Templeton Foundation, the University of California, and the National Science Foundation under Grant No. NSF PHY11-25915.

\appendix
 
\section{Derivation of the composite vortex liquid boundary lagrangian}
\label{CVLboundaryderivation}
 
We describe the bulk physics of the CVL with the lagrangian,
\begin{align}
\label{CVLbulkappendix}
{\cal L}_{\rm CVL} = \psi^\dagger_{\rm v} \Big(i \partial_t + e_\ast b_t + {1 \over 2 m_{\rm v}} (\partial_j - i e_\ast b_j )^2  \Big) \psi_{\rm v} - {e_\ast^2 \over 4 \pi} \epsilon^{\mu \nu \rho} \Big(b_\mu \partial_\nu b_\rho - 2 \tilde{A}_\mu \partial_\nu (A_\rho - b_\rho) + \tilde{A}_\mu \partial_\nu \tilde{A}_\rho \Big).
\end{align}
If the CVL is placed on the lower half-plane ($y<0$) (or any space with boundary, more generally) with topologically trivial vacuum in the upper half-plane ($y>0$), boundary degrees of freedom living at $y=0$ are required to maintain gauge invariance.
Our task in this appendix is to derive the boundary lagrangian governing their dynamics.

The $\tilde{A}_t = 0$ gauge imposes the constraint,
\begin{align}
\tilde{A}_i = A_i - b_i + \partial_i \phi_1, 
\end{align}
where $\phi_1 \mapsto \phi_1 - (\Lambda_{A} - \Lambda_{b})$ under the gauge transformations $A_i \mapsto A_i + \partial_i \Lambda_A$ and $b_i \mapsto b_i + \Lambda_{b}$. 
We choose the gauge variation of $\tilde{A}_\mu$ to vanish on the boundary at $y=0$.
Substituting the solution of the constraint into ${\cal L}_{\rm CVL}$, the bulk lagrangian becomes
\begin{align}
\label{CVLbulkstepone}
{\cal L}_{\rm CVL} = \psi^\dagger_{\rm v} \Big(i \partial_t + e_\ast b_t + {1 \over 2 m_{\rm v}} (\partial_j - i e_\ast b_j )^2  \Big) \psi_{\rm v} - {e_\ast^2 \over 4 \pi} \epsilon^{\mu \nu \rho} \Big(b_\mu \partial_\nu b_\rho - (b_\mu - A_\mu) \partial_\nu (b_\rho - A_\rho)\Big).
\end{align}
In addition, we obtain a contribution to the boundary lagrangian at $y=0$,
\begin{align}
\label{boundarycontribution}
{\cal L}_{\partial{\rm CVL}} \supset {e_\ast^2 \over 4 \pi} \Big[(\partial_t \phi_1 + A_t - b_t) (\partial_x \phi_1 + A_x - b_x) - v_{1}(\partial_x \phi_1 + A_x - b_x)^2 + \epsilon_{\mu \nu y} (A_\mu - b_\mu) \partial_\nu \phi_1 \Big],
\end{align}
after performing an integration by parts to put ${\cal L}_{\rm CVL}$ into the above form.
A second boundary degree of freedom of opposite chirality to $\phi_1$ is required to ensure invariance under the transformation $b_\mu \mapsto b_\mu + \Lambda_{b}$.
Because the coefficient of the Chern-Simons term proportional to $\epsilon^{\mu \nu \rho} b_\mu \partial_\nu b_\rho$ vanishes in Eq. (\ref{CVLbulkstepone}), the anomalous gauge variation proportional to $\Lambda_{b} \epsilon_{\mu \nu y} \partial_\mu b_\nu$ of the lagrangian governing $\phi_1$ can be entirely canceled by the variation of the lagrangian for the boundary degree of freedom represented by $\phi_2$ which transforms as $\phi_2 \mapsto \phi_2 + \Lambda_{b}$.
(Recall that the anomalous variation appears at the classical level in this bosonized description -- see, for example, [\onlinecite{Mulligantheta2011}].)
Thus, we find the boundary lagrangian,
\begin{align}
\label{CVLboundary}
{\cal L}_{\partial {\rm CVL}} & = {e_\ast^2 \over 4 \pi} \Big[(\partial_t \phi_1 + A_t - b_t) (\partial_x \phi_1 + A_x - b_x) - v_{1}(\partial_x \phi_1 + A_x - b_x)^2 + \epsilon_{\mu \nu y} (A_\mu - b_\mu) \partial_\nu \phi_1 \cr
& - (\partial_t \phi_2 - b_t) (\partial_x \phi_2 - b_x) - v_{2}(\partial_x \phi_2 - b_x)^2 + \epsilon_{\mu \nu y} b_\mu \partial_\nu \phi_2\Big]\delta(y=0).
\end{align}
The terms quadratic in spatial derivatives, which represent intra-boundary mode density-density interactions, have been added by hand and are parameterized by the velocities $v_1, v_2 > 0$.

\section{Non-relativistic limit of the particle-vortex symmetric liquid}
\label{nonrellimit}

In this appendix, we review the non-relativistic limit of the particle-vortex symmetric liquid deformed by a finite Dirac mass $m \bar{\psi} \psi$.
Restoring the ``speed of light" $c$, i.e., setting $t \rightarrow c t$, $\alpha_t \rightarrow {1 \over c} \alpha_t$, and $A_t \rightarrow {1 \over c} A_t$, the deformed particle-vortex symmetric liquid lagrangian becomes
\begin{align}
\label{pvsymappendix}
{\cal L}_{\rm pv} = {1 \over c} \bar{\psi} i \gamma^t D_t \psi + \bar{\psi} i \gamma^j D_j \psi + m c \bar{\psi} \psi + {e_\ast^2 \over 4 \pi c} \epsilon^{\mu \nu \rho} \Big( {1 \over 2} \alpha_\mu \partial_\nu \alpha_\rho - 2 A_\mu \partial_\nu \alpha_\rho + A_\mu \partial_\nu A_\rho \Big),
\end{align}
where $D_\mu = \partial_\mu - i e_\ast \alpha_\mu$.

At finite external magnetic field $\partial_x A_y - \partial_y A_x \equiv B > 0$, the anti-particles (or holes) of $\psi$ are ``massive." 
Focusing on the ``light" particles, we write 
\begin{align}
\psi = e^{- i |m| c^2 t} \begin{pmatrix} \psi_{\rm v} \cr \psi_\Phi \end{pmatrix}
\end{align}
and substitute into the lagrangian to find
\begin{align}
\label{pvsymappendix}
{\cal L}_{\rm pv} & = c \Big(|m| + m\Big) \psi^\dagger_{\rm v} \psi_{\rm v} + c \Big(|m| - m\Big) \psi^\dagger_\Phi \psi_\Phi + \psi_{\rm v}^\dagger i D_{\bar{z}} \psi_\Phi + \psi^\dagger_\Phi i D_z \psi_{\rm v} \cr
& + {1 \over c} \Big(\psi_{\rm v}^\dagger D_t \psi_{\rm v} + \psi_\Phi^\dagger D_t \psi_\Phi\Big) + {e_\ast^2 \over 4 \pi c} \epsilon^{\mu \nu \rho} \Big( {1 \over 2} \alpha_\mu \partial_\nu \alpha_\rho - 2 A_\mu \partial_\nu \alpha_\rho + A_\mu \partial_\nu A_\rho \Big),
\end{align}
where $D_- \equiv D_y - i D_x$ and $D_+ \equiv D_y + i D_x$.
For $m > 0$, the $\psi_{\rm v}^\dagger$ equation of motion sets
\begin{align}
\psi_{\rm v} = - {i \over 2 m c} D_+ \psi_\Phi
\end{align}
which to leading order in a $1/c$ expansion gives
\begin{align}
{\cal L}_{\Phi} = {1 \over c} \psi_\Phi^\dagger D_t \psi_\Phi + {1 \over 2 m c} \psi^\dagger_\Phi \Big(D_- D_+ \Big) \Psi_\Phi + {e_\ast^2 \over 4 \pi c} \epsilon^{\mu \nu \rho} \Big( {1 \over 2} \alpha_\mu \partial_\nu \alpha_\rho - 2 A_\mu \partial_\nu \alpha_\rho + A_\mu \partial_\nu A_\rho \Big).
\end{align}
Alternatively, for $m < 0$, we use the $\psi^\dagger_\Phi$ equation of motion to solve for $\psi_\Phi$ in terms of $\psi_{\rm v}$.
When substituted into ${\cal L}_{\rm pv}$, we obtain 
\begin{align}
{\cal L}_{\rm v} = {1 \over c} \psi_{\rm v}^\dagger D_t \psi_{\rm v} + {1 \over 2 m c} \psi^\dagger_{\rm v} \Big(D_+ D_- \Big) \Psi_{\rm v} + {e_\ast^2 \over 4 \pi c} \epsilon^{\mu \nu \rho} \Big( {1 \over 2} \alpha_\mu \partial_\nu \alpha_\rho - 2 A_\mu \partial_\nu \alpha_\rho + A_\mu \partial_\nu A_\rho \Big).
\end{align}
Up to a positive (negative) shift of the coefficient of the Chern-Simons term for $\alpha_\mu$ by ${e_\ast^2 \over 8 \pi c}$ and the ``Zeeman" couplings  $- {e_\ast \over 2 m c} (\partial_x \alpha_y - \partial_y \alpha_x) \psi_\Phi^\dagger \psi_\Phi$ and $ {e_\ast \over 2 m c} (\partial_x \alpha_y - \partial_y \alpha_x) \psi_{\rm v}^\dagger \psi_{\rm v}$, ${\cal L}_\Phi$ and ${\cal L}_{\rm v}$ are identical to the CBL and CVL lagrangians.
The Zeeman couplings are allowed by symmetry, but are of higher order in the derivative expansion of the effective lagrangians.

Evidently, the ``massive" anti-particles contribute a Chern-Simons term for $\alpha_\mu$ with coefficient equal to ${e_\ast^2 \over 8 \pi} {m \over |m|}$.
This may be seen without explicit calculation simply from the requirement of the preservation of the ``ultraviolet" and ``infrared" contributions to the parity anomaly constraint [\onlinecite{NiemiRQ, Redlichparitylong, AlvarezGaumeWitten1984}]: the self-dual dyon fermion determinant contribution to the would-be anomaly must be matched by that of the infrared theory.  
Under the assumption that the non-relativistic fermions in ${\cal L}_{\Phi}$ and ${\cal L}_{\rm v}$ make no contribution, the desired Chern-Simons term must be generated upon decoupling the massive anti-particle.
Adding the anti-particle contribution to ${\cal L}_\Phi$ or ${\cal L}_{\rm v}$, we recover the CBL or CVL lagrangians.


\section{Wiedemann-Franz}
\label{WF}

Inspired by the observation in [\onlinecite{WangSenthilsecond2016}, \onlinecite{WangSenthil2016}], we consider the possible violation of the Wiedemann-Franz ``law."
Under the assumption that there exists a Wiedmann-Franz law between the electronic contribution to the thermal conductivity $\kappa_{xx}$
and self-dual dyon conductivity $\sigma_{xx}^\psi$ with Lorenz number ${\cal L}$ equal to that of a Fermi liquid of charge $e_\ast$ electrons, we consider the ratio,
\begin{align}
\label{ratio}
r_L = {1 \over {\cal L}} {\kappa_{xx} \over T \sigma_{xx}}.
\end{align}
$\sigma_{xx}$ is the measured longitudinal electrical conductivity.
The possibility of a Wiedemann-Franz ``law" is an immediate consequence of scaling about the pertinent low-energy fixed point; its realization requires that the thermal and electrical current relaxation mechanisms be the same (or at least the operators which relax the currents have equal scaling dimensions).
In assuming a Wiedemann-Franz ``law" between the thermal conductivity and self-dual dyon conductivity, we are ignoring any contribution to the thermal conductivity from the emergent gauge fields.

The ratio $r_L$ quantifies the possible deviation from the behavior expected from a Fermi liquid.
This Fermi liquid may be thought of as the ``normal state" electrons extrapolated to $T \rightarrow 0$. 
Using Eq. (\ref{response}), we can express $\sigma_{xx}^\psi$ in terms of the measured electrical resistivity to find:
\begin{align}
r_L & = {\rho_{xx}^2 + \rho_{xy}^2 \over \rho_{xx}^2 + ({2 \pi \over e_\ast^2} + \rho_{xy})^2} \cr
& \approx {1 \over 2 (1 + {e_\ast^2 \over 2 \pi} \rho_{xy})}.
\end{align}
The second equality obtains for approximate self-duality.
Near a SIT with approximate self-duality with $\rho_{xx} \approx h/4e^2$ and $\rho_{xy} \approx 0$ [\onlinecite{Hebard1990, PaalanenHebardRuel1992,Yazdani1995, Steiner2008, Ovadia:2013aa, Breznay2016}], $r_L \approx 1/2$.
Deviations away from self-duality with $\rho_{xy} = 0$ and $\rho_{xx} \ll 2\pi/e_\ast^2$, result in $r_L < 1/2$.

This behavior contrasts that which is expected [\onlinecite{WangSenthilsecond2016}] using a composite fermion treatment for the half-filled Landau level where the analog of the factor $({2 \pi \over e_\ast^2} + \rho_{xy}) \equiv ({h \over 4 e^2} + \rho_{xy})$ is replaced by $({2 h \over e^2} + \rho_{xy}) \approx 0$ and $\rho_{xx} \ll \rho_{xy}$ so that $r_L \gg 1$.
As the strength of the disorder is increased so that the metallic phase pinches into an IQHT critical point, $\rho_{xx}$ is increased and $r_L \rightarrow 1$ similar to a self-dual SIT.



\bibliography{emergentpv}{}

\end{document}